\documentclass[12pt]{article}
\usepackage{amsmath,amssymb,array,cite,latexsym, arydshln,marvosym,graphicx}
\usepackage{caption,subcaption}
\usepackage[usenames,dvipsnames]{color}
\usepackage{authblk}
\usepackage{soul}

\newcommand{\be}{\begin{equation}}
\newcommand{\ee}{\end{equation}}
\newcommand{\bea}{\begin{eqnarray}}
\newcommand{\eea}{\end{eqnarray}}

\title{Leaky modes of waveguides as a classical optics analogy of quantum resonances}

\author[1]{Sara Cruz~y~Cruz}
\author[2]{Oscar Rosas-Ortiz}

\affil[1]{\footnotesize UPIITA, Instituto Polit\'ecnico Nacional, Av. IPN No. 2580, Col. La Laguna Ticom\'an, C.P. 07340 M\'exico~D.F. Mexico} 

\affil[2]{\footnotesize Physics Department, Cinvestav, AP 14-740, 07000 M\'exico~D.F., Mexico}

\date{}
\begin{document}
\maketitle

\begin{abstract}{\footnotesize
A classical optics waveguide structure is proposed to simulate resonances of short range one-dimensional potentials in quantum mechanics. The analogy is based on the well known resemblance between the guided and radiation modes of a waveguide with the bound and scattering states of a quantum well. As resonances are scattering states that spend some time in the zone of influence of the scatterer, we associate them with the leaky modes of a waveguide, the latter characterized by suffering attenuation in the direction of propagation but increasing exponentially in the transverse directions. The resemblance is complete since
resonances (leaky modes) can be interpreted as bound states (guided modes) with definite lifetime (longitudinal shift). As an immediate application we calculate the leaky modes (resonances) associated with a dielectric homogeneous slab (square well potential) and show that these modes are attenuated as they propagate.}
\end{abstract}


\section{Introduction}

In general, a {\em resonance} may be defined as the excitation of a system that results when one of its characteristic frequencies matches a particular value that is defined by either the boundary conditions, the external forces, or any other interaction or constraint applied to the system. The concept arises from the study of oscillating systems in classical mechanics and extends its applications to physical theories like electromagnetism, optics, acoustics, and quantum mechanics, among others \cite{Ros08}. In quantum and nuclear physics the resonances are associated with metastable states of a system which has sufficient energy to break up into two or more subsystems \cite{Moi98}. The whole range of scattering experiments provides a big number of examples in this matter \cite{Tay72,Bra89}. This last is because scattering includes the situation in which the impinging particle is `captured' for a while in the zone of influence of the scatterer,  so that the projectile and the scatterer form a new unstable system (resonance state). The capturing occurs for specific energies of the projectile that are defined by the general properties of the scatterer. Eventually, the projectile is released and escapes from the interaction zone (decay process). In this model the `time of capture' corresponds to the lifetime of the decaying system that is formed of the scatterer and the projectile under the resonance condition. Collisions, in addition, are modeled as interactions localized in time and space. This implies that the involved potential vanishes rapidly enough in space, so that incoming and outgoing asymptotic states can be represented by wave packets in free motion. The problem is usually reduced to the analysis of one-dimensional effective potentials such that the main information is obtained from the transmission and reflection amplitudes of the scattering states, see e.g. \cite{Nus59}. In the case of one-dimensional square potential wells and barriers the resonances appear for specific values of the parameters that define the interaction. Indeed, for such parameters the resonances can be associated with the bell-shaped peaks of the corresponding transmission coefficient \cite{Fer08a,Klai10}. The presence of resonances can be also stimulated either by shallowing a square well \cite{Zav04} or by adding to it a static electric field \cite{Emm05}. As resonances are special cases of scattering states, they are represented by irregular (not normalizable) vectors \cite{del02,del04} that satisfy the purely outgoing condition \cite{Sie39} (see also \cite{Ros08}). This last property of resonances is useful in constructing complex potentials that operate as optical devices with position-dependent complex refractive index \cite{Fer08a,Fer07,Fer08b}.

Remarkably, quantum wells and barriers can be made with high precision and compositional control by modern epitaxial growth techniques \cite{Har09}. This has opened a wide window of experimental possibilities to explore various quantum phenomena in real devices and actual laboratory settings. For example, it has been shown that the transmission time of a wave packet that is tunnelling through a one-dimensional barrier does not depend on the barrier thickness \cite{Har62}; a similar property is found in one-dimensional potential wells \cite{Li00,Vet01}. This condition implies anomalously large group velocities of the wave packets that impinge the barrier (well) since the velocity of the packet's peak must increase with the width across the interaction zone. Such a superluminal phenomenon has been observed using electromagnetic analogues for evanescent modes \cite{Mar92} and microwave pulses \cite{End92}, at the same time this result has stimulated the designing of high-speed devices based on the tunneling properties of semiconductors (see, e.g., Chs. 11 and 12 of Ref. \cite{Mug08}). On the other hand, as the scattering by quantum wells attenuates the outgoing wave packets only because of the multiple reflections at the well boundaries, it is expected that the resonance condition implies a maximum in the time spend by the projectile in traversing the zone of influence of the scatterer \cite{Fer08a}. Recent theoretical research includes the case of a rectangular well that is embedded in an environment formed of a zero potential energy (flat potential) at the right and a parabolic potential at the left of the well \cite{Fer11,Ros12,Ros13}. Thus, the connection between time delay and `time of capture' associated to the scattering processes would be verified in the laboratory by using semiconductor materials.

Another accessible laboratory tool to study quantum phenomena is provided by waveguide structures. The connection is based on the analogies between light waves and Schr\"odinger wave-state functions that have been studied since the early years of quantum physics \cite{Sla28}. In this approach the amplitude of an electromagnetic field is associated to the probability amplitude of a quantum state with two degrees of freedom while the propagation of such field through the waveguide corresponds to the time-evolution of the quantum state. Such optical-quantum analogy is representative of a wider scheme that includes diverse connections between classical and quantum systems \cite{Dra04}, and has been successfully applied in studying Anderson localization \cite{Sch07}, coherent enhancement \cite{Vor03} and destruction of tunneling \cite{Del07}, coherent population transfer \cite{Kar07}, and decay of metastable states coupled to semi-infinite, tight-binding lattices \cite{Lon06}, among others. Of special interest for the purposes of this paper, nanoseconds delay-time measurements have been performed into the electromagnetic framework by using microwave setups of narrowed waveguides \cite{Ran91}. The relevance of this last result obeys the fact that the waveguide simulates a quantum potential barrier so that the measured electromagnetic delay-time is in correspondence with the quantum tunneling time. Thus, the waveguide structures represent an emergent and versatile arena to perform realistic experiments addressed to get deeper insights in the understanding of resonances in quantum physics.

The aim of this paper is to get a classical optics analogy of the resonances associated to one-dimensional short range potentials in quantum physics. The centerpiece of our approach is the connection between the mathematical structure of the propagation of electromagnetic signals in waveguides and the dynamics of a time-dependent quantum state. In this form our interest is twofold: we want to connect the theoretical calculations of resonances with mensurable properties of an electromagnetic wave that propagates in a waveguide, and we would like to extend the application of the resonance mathematical techniques to the solving of problems associated to electromagnetic waveguides. A collateral result will be the connection between the resonance properties of electrons that are injected into heterostructures and the properties of the electromagnetic signals that propagate in optical waveguides. The paper is organized as follows. In Section~\ref{resonances} we revisit the main properties of the resonance states associated to one-dimensional short range potentials. In particular, it is shown the relevance of the Fock-Breit-Wigner distribution in a scattering process. When the transition amplitude from an initial scattering state to its time-evolved version is ruled by such a distribution one immediately identifies energy resonances between the projectile and the scatterer. It is also shown that analytical continuation of the eigenvalue equation to a complex plane of energies is required to get an appropriate mathematical description of resonance states. In Section~\ref{modes} we include some of the well known generalities of the mode fields in waveguide structures and their connection with the Schr\"odinger equation of a system of two degrees of freedom. The case of electromagnetic propagation through a finite, dielectric homogeneous material and its association with a one-dimensional square well potential is considered in Section~\ref{well} as immediate application. Section~\ref{leaky} is devoted to the connection between resonance states and a singular form of modal field in waveguides, the latter is known as {\em leaky mode} and is characterized by suffering attenuation in the direction of propagation but increasing exponentially in the transverse directions. We apply the resonance approach of Ref.~\cite{Fer08a} in the calculation of the leaky modes associated to the dielectric homogeneous slab described in Section~\ref{well}. As an example we provide the eigenvalues (resonances) corresponding to the leaky modes supported by a slab of definite width and refractive index, and show that these modes are attenuated as they propagate. Section~\ref{shift} includes the calculation of the longitudinal shifts for the leaky modes of Section~\ref{leaky}, we show that the maxima of these shifts are in correspondence with the maxima of the transmission coefficient, a result that verifies the connection between times of capture and resonances in the scattering processes. Finally, Section~\ref{concluye} includes some of the main conclusions of the present work.

\section{One-dimensional scattering states and resonances}
\label{resonances}

Let $H$ be the Hamiltonian defined by a short range one-dimensional potential $U(x)$. Assuming that the spectrum of $H$ is formed of discrete and continuous eigenvalues one can consider the following sets of state-vectors
\be
\mbox{discrete eigenvalues} \, E_k \, \leftrightarrow \,  \left\{
\begin{array}{l}
\vert \phi_k \rangle, \, k\in {\cal I}_D \subseteq \mathbb Z^+\\[1ex]
\mathbb I_D = \sum_{k \in \cal I_D} \vert \phi_k \rangle \langle \phi_k \vert\\[1ex]
{\cal H}_D = \mbox{Span} \{ \vert \phi_k \rangle \}_{k \in \cal I_D}
\end{array}
\right.
\label{basis1}
\ee

\be
\mbox{continuous eigenvalues} \, E \, \leftrightarrow \, \left\{
\begin{array}{l}
\vert \phi_E \rangle, \, E \in {\cal I}_E \subseteq \mathbb R\\[1ex]
\mathbb I_E = \int_{E \in {\cal I}_E} dE \vert \phi_E \rangle \langle \phi_E \vert\\[1ex]
{\cal H}_E = \mbox{Span} \{ \vert \psi_E \rangle \}_{E \in {\cal I}_E}
\end{array}
\right.
\label{basis2}
\ee
where $\mathbb I_D$ and $\mathbb I_E$ are respectively the identity operator in ${\cal H}_D$ and ${\cal H}_E$. Consistently, the spectrum and Hilbert space of $H$ are  $\sigma(H) = \{E_k \}_{k \in \cal I_D} \cup \{ E \}_{E \in {\cal I}_E}$ and ${\cal H} ={\cal H}_D \oplus {\cal H}_E$ respectively. As usual, the discrete and continuous eigenvectors (\ref{basis1})--(\ref{basis2}) are orthonormal in their respective subspace according to the rules
\be
\langle \phi_j \vert \phi_k \rangle = \delta_{jk}, \quad \langle \phi_E \vert \phi_{E'} \rangle = \delta (E-E').
\label{ortho}
\ee
Hereafter we pay detailed attention to the scattering states (i.e., vectors in ${\cal H}_E$). The description of bound states (i.e., vectors in ${\cal H}_D$) is straightforward by using the well known relationship $\int dE \leftrightarrow \sum_k$. The reason is that, as we have said in the introduction, resonance states arise quite naturally in the scattering processes.

Any scattering state $\vert \psi \rangle \in {\cal H}_E$ can be expressed in the basis of continuous eigenvectors
\be
\vert \psi \rangle = \mathbb I_E \vert \psi \rangle = \int_{E \in {\cal I}_E} dE C(E) \vert \phi_E \rangle, \quad C(E) = \langle \phi_E \vert \psi \rangle,
\label{sca1}
\ee
and is such that
\be
\vert \vert \psi \vert \vert^2= \langle \psi \vert \psi \rangle = \int_{E \in {\cal I}_E} dE \vert C(E) \vert^2.
\label{sca2}
\ee
In $x$-representation the above expressions read as
\be
\psi(x) = \int_{E \in {\cal I}_E} dE C(E) \phi_E(x), \quad C(E) = \int_{\mathbb R} dx \overline \phi_E(x) \psi(x),
\label{sca3}
\ee
and
\be
\vert \vert \psi \vert \vert^2= \int_{\mathbb R} dx \vert \psi(x) \vert^2= \int_{E \in {\cal I}_E} dE \vert C(E) \vert^2,
\label{sca4}
\ee
with $\overline z$ standing for the complex conjugate of $z\in \mathbb C$. The above expressions allow to write the energy distribution $\omega(E)$ as follows
\be
\omega(E) = \frac{dW}{dE} = \vert C(E) \vert^2.
\label{ener1}
\ee
Let (\ref{sca1}), equivalently (\ref{sca3}), be the initial state of a stationary system. The transition amplitude $T(t\geq 0)$ from $\vert \psi (t=0)\rangle = \vert \psi_0 \rangle$ to $\vert \psi(t)\rangle = \vert \psi_t \rangle$ is given by the inner product
\be
T(t\geq0)= \langle \psi_0 \vert \psi_t \rangle= \int_{\mathbb R} dx \overline \psi_0 (x) \psi_t(x) = \int_{E \in {\cal I}_E} dE \omega(E) e^{-iEt/\hbar}.
\label{prob1}
\ee
It is clear that $T$ can be investigated in terms of spatial coordinates $x$ or as a function of the energy distribution (\ref{ener1}). In this work we shall assume that $\omega(E)$ is given by the following expression
\be
\omega(E) =
\frac{(\Gamma/2)^2}{\left(E-E_0 +i\Gamma/2 \right)\left(E-E_0 -i\Gamma/2 \right)}=
\frac{(\Gamma/2)^2}{\left(E-E_0\right)^2 +\left(\Gamma/2 \right)^2},
\label{fock1}
\ee
where $E_0 \in {\cal I}_E$ is the energy of the initial state $\vert \psi_0 \rangle$ and $\Gamma \geq 0$ is a parameter having units of energy. Note that the bell-shaped function (\ref{fock1}), it is centered at $E=E_0$, for which $\omega(E_0)=1$, and has a half-width at half-maximum equal to $\Gamma/2$. In addition, $\omega(E)$ is a meromorphic function that has two poles $E_{\pm}= E_0 \pm i\Gamma/2$. This function is known as either Cauchy (mathematics), Lorentz (statistical physics) or Fock-Breit-Wigner -FBW- (nuclear and particle physics) distribution. Using (\ref{ener1}) we can write the Fourier coefficient $C(E)$ in the expansion (\ref{sca1}) as follows
\be
C(E)= \frac{\Gamma/2}{E-E_0 +i\Gamma/2}.
\label{Four1}
\ee
After the analytic continuation of ${\cal I}_E \subseteq \mathbb R$ to the complex $E$-plane one can introduce (\ref{fock1}) into (\ref{prob1}) to arrive, up to a global constant factor, at the expression
\be
T(t\geq0)= \frac{\Gamma}{2} e^{-i \varepsilon_0 t/\hbar} = \left( \frac{\Gamma}{2} e^{-i E_0 t/\hbar} \right) e^{-\Gamma t /(2\hbar)}
, \quad \varepsilon_0:= E_0 - i\frac{\Gamma}{2}.
\label{prob2}
\ee
This last result has the form of a transient oscillation \cite{Ros08} and means that the transition from $\vert \psi_0 \rangle$ to $\vert \psi_t \rangle$ is an exponential decreasing function of time whenever $C(E)$ is given by Eq.~(\ref{Four1}). That is, the initial state $\vert \psi_0 \rangle$ decays according to the time-dependent probability
\be
\vert T (t\geq 0) \vert^2 = \left( \frac{\Gamma}{2} \right)^2 e^{-\Gamma t /\hbar}.
\label{prob3}
\ee
As we can see, the width $\Gamma$ of the distribution (\ref{fock1}) represents an indirect measure of the lifetime of $\vert \psi_0 \rangle$ because at time $\tau := \hbar/\Gamma$ the probability that this state has not yet decayed is reduced approximately in $36\%$, i.e., $\Gamma^2/4 \rightarrow \Gamma^2/(4e)$. It is clear that $\Gamma$ and $\tau$ are correlated, the smaller the value of $\Gamma$ the larger the lifetime $\tau$. In general, one says that there is a resonance $E_0$ of width $\Gamma$ when either the energy distribution of the system at time $t\geq 0$ is given by the FBW function (\ref{fock1}) or the Fourier coefficients $C(E)$ of the expansion (\ref{sca1}) are given by (\ref{Four1}).

We would like to emphasize that the dependence of $\omega$, $C$ and $T$ on the  complex number $\varepsilon_0 = E_0 -i \Gamma/2$, as this has been expressed in Eqs.~(\ref{fock1})--(\ref{prob3}), is not merely aesthetic; this corresponds to the formulation of Gamow, Fock, Breit and Wigner in the late 1920s and along the 1930 decade (see details in \cite{Ros08}). It is due to Siegert \cite{Sie39} that a resonance is introduced as a solution $u_{\varepsilon}$ of the Schr\"odinger equation associated to the complex eigenvalue $\varepsilon = E-i \Gamma/2$ and satisfying the purely outgoing condition \cite{Fer08a}:
\be
\lim_{x  \rightarrow \pm \infty}\beta = \mp ik,
\label{pure1}
\ee
where $k^2 = 2m\varepsilon /\hbar^2 \in \mathbb C$, and
\be
\beta:= -\frac{d}{dx} \ln u_{\varepsilon}.
\label{pure2}
\ee
It can be shown that the above conditions lead automatically to the FBW distribution (\ref{fock1}) with $k$ a number in the fourth quadrant of the complex plane $\mathbb C$, see \cite{Ros08}. In turn, the distribution of the complex points $\varepsilon^{1/2}=\hbar k/\sqrt{2m}$ characterizes a Riemman surface of two sheets, each one cutting along the positive real axis. These complex numbers define directly the behavior of the propagator (\ref{prob3}). For instance, consider a very narrow distribution $\omega(E)$, that is $\Gamma <<1$ (the point $\varepsilon \in \mathbb C$ is close to the positive real axis). As  $\Gamma$ defines the vicinity $E \in (E_0 -\Gamma/2, E_0 + \Gamma/2)$, we see that `large lifetimes' $\tau = \hbar/\Gamma$ means narrow widths $\Gamma$, so that the energy $E$ of the system at time $t\geq 0$ is very close to the initial energy $E_0$. In the extremal case $\Gamma \rightarrow 0$, the distribution $\omega (E)$ is as narrow and singular as $\delta(E-E_0)$. That is, the related lifetime $\tau$ is large enough $(\tau \rightarrow +\infty)$ to consider the initial vector $\vert \psi_0 \rangle$ a non-decaying (stable) state. In this context, bound states are considered as decaying states with infinitely large lifetimes and vice versa, unstable (decaying) states correspond to bound states with a definite lifetime.

To get more insights about the complex eigenvalue $\varepsilon_0$ let us consider the Schr\"odinger equation
\be
H  \psi(x; t)  = i \hbar \frac{d}{dt}  \psi(x;t).
\label{schro1}
\ee
Assuming separation of variables $\psi(x;t) = \varphi(x) \theta(t)$, with $\theta(t) = e^{-iEt/\hbar}$ and $E \in {\cal I}_E$ the separation constant, the probability density does not depend on time $\vert \psi(x; t) \vert^2 = \vert \varphi(x) \vert^2$. Now, let us make the analytic continuation of the eigenvalue equation $H \varphi(x) = E \varphi(x)$ in order to include complex eigenvalues $\varepsilon = E -i\Gamma/2$, that is $Hu_{\varepsilon}= \varepsilon u_{\varepsilon}$. In this case we have $\psi(x;t) = u_{\varepsilon}(x) e^{-i\varepsilon t/\hbar}$, and the probability density is an exponential decreasing function of time $\vert \psi(x;t) \vert^2 = \vert u_{\varepsilon}(x) \vert^2 e^{-\Gamma t/\hbar}$. Clearly, the spatial part $u_{\varepsilon}(x)$ of the wave-function  $\psi(x;t) $ is no longer a square-integrable function because this belongs to the complex eigenvalue $\varepsilon$ and the Hamiltonian $H$ is still Hermitian. Indeed, it can be proved that $u_{\varepsilon}(x)$ diverges exponentially as $\vert x \vert \rightarrow +\infty$. Thus, the density $\vert \psi(x; t) \vert^2$ increases exponentially for either large $\vert x \vert$ or large negative values of $t$. The usual form to avoid some of the complications connected with the limit $t\rightarrow -\infty$ is to consider the long lifetime limit $\Gamma \rightarrow 0$, see e.g. \cite{Boh99}. Therefore, we consider the condition
\be
\frac{\Gamma/2}{\Delta E} <<1.
\label{cond1}
\ee
This last means that the level width $\Gamma$ must be much smaller than the level spacing $\Delta E$ in such a way that closer resonances imply narrower widths (longer lifetimes). In general, the main difficulty is precisely to find the adequate $E$ and $\Gamma$. However, it has been shown \cite{Fer08a} that for one-dimensional stationary short range potentials the superposition of a denumerable set of FBW distributions (each one centered at each resonance $E_n$, $n=1,2\ldots$) entails an approximation of the transmission coefficient $T$ such that the larger the number $N$ of  resonances involved, the higher the precision of the approximation:
\be
T \approx \omega_N(E) = \sum_{n=1}^N \omega(E,E_n); \qquad \omega(E,E_n) = \frac{(\Gamma_n/2)^2}{(E-E_n)^2 +(\Gamma_n/2)^2}.
\label{FBW1}
\ee

\section{Mode fields in waveguide structures}
\label{modes}

Consider the Helmholtz equation
\be
\frac{\partial^2}{\partial x^2}E(x,z) + \frac{\partial^2}{\partial z^2}E(x,z) + k_0^2 n^2(x) E(x,z) =0,
\label{Hel1}
\ee
with $k_0=w/c$ the wave number in vacuum. Equation (\ref{Hel1}) can be achieved from the Maxwell equations for transversal electric fields $\vec E$ in presence of inhomogeneous dielectric materials, neglecting the magnetic polarizability and the dispersive properties of the medium. Indeed, (\ref{Hel1}) corresponds to an electric wave propagating along the positive $z$-direction and polarized in the $y$-direction, assuming that the refractive index $n$ depends only on the $x$-coordinate. In the paraxial regime we use the ansatz
\be
E(x,z) = \varphi(x) e^{-ik_0(\gamma -n_0)z},
\label{ans}
\ee
to transform (\ref{Hel1}) into the linear second order differential equation
\be
\left[-\frac{1}{2k_0^2n_0} \frac{d^2}{dx^2} - n (x) \right] \varphi (x)= \varepsilon \varphi (x), \quad \varepsilon = \gamma - n_0,
\label{Hel2}
\ee
known as the paraxial Helmholtz equation. In (\ref{ans})--(\ref{Hel2}) the number $n_0$ is a reference refractive index, that may be taken as the maximum of $n(x)$, and we have assumed that the weakly guiding condition ($\Delta n = \vert n-n_0 \vert << 1$) holds. In this approach the parameter $\varepsilon$, known as the propagation constant, is the value of the $z$-component of the generalized linear momentum \cite{Glo69} and defines the slope of the light beam as follows
\be
\varepsilon =- p_z = -n(x)\cos\theta(x).
\label{pz}
\ee

In general, modal fields associated to open waveguides are classified into guided (trapped) modes and radiation modes, the latter propagating outside the core \cite{Sny74,Sny83,Hu09,Tre14}. The guided modes are bounded fields that correspond to real values of the propagation constant fulfilling $-n_0 \leq \varepsilon < -1$. They are perfectly guided through the waveguide as they suffer no  attenuation along the optical axis and decay exponentially in the transverse direction. Thus, as guided modes are trapped into the core of the waveguide they add no energy flux into the infinite surrounding clad. In turn, propagating radiation modes correspond to real propagation constants fulfilling $-1 \leq \varepsilon < 0$, so that they are bounded oscillatory fields in the transverse as well as in the longitudinal directions \cite{Sny74}. They appear as a consequence of the infinite extension of the clad cross-section. Guided  and radiation modes constitute a complete set of modal fields in the sense that any physical solution to the paraxial Helmholtz equation can be expressed as a linear superposition of a finite number of guided modes and a continuum of radiation modes.

At this stage it is important to stress the resemblance of the electric field expression (\ref{ans}) with the solutions $\psi(x;t) = \varphi(x) e^{-iEt/\hbar}$ of the Schr\"odinger equation (\ref{schro1}). This allows the identification
\be
k_0^{-1} \leftrightarrow \hbar, \quad z \leftrightarrow t, \quad \varepsilon = \gamma-n_0 \leftrightarrow E.
\label{ident1}
\ee
In adittion, Eq.~(\ref{Hel2}) would correspond to the eigenvalue equation $H\varphi = E\varphi$, provided the following additional identification holds
\be
n_0 \leftrightarrow m, \quad -n(x) \leftrightarrow U(x), \quad H = -\left(\frac{k_0^{-2}}{2 n_0} \right) \frac{d^2}{dx^2} - n(x).
\label{ident2}
\ee
The identification (\ref{ident1})--(\ref{ident2}) is well known in the literature, see e.g. \cite{Glo69}. This allows to construct the classical optics analogy of diverse quantum phenomena and vice versa, the quantum analogy of classical optical phenomena \cite{Glo69,Dra04}. The analogy includes the sets of vectors introduced in (\ref{basis1}) and (\ref{basis2}) because bound states correspond to `trapped' waves that have no chance of escaping from the zone of influence of the scatterer (as we have seen in the previous section, these states can be interpreted as decaying states with infinitely large lifetime). Thus, quantum bound states correspond to guided electromagnetic modes. Particles in scattering states, on the other hand, can be temporally trapped by the scatterer in such a form that their presence in the zone of interaction decreases in time. In this form, scattering states correspond to radiation modes. The orthonormalization properties of guided and radiation modal fields can be stated in a form similar to that of the quantum mechanical bound and scattering states (see, e.g. \cite{Hu09,Sny74}). Hence, the total radiated power can be expressed as the sum of two terms: a first one associated to the power that propagates inside the core, given by a finite sum containing the contribution of guided modes, and a second term corresponding to the power radiated from the core, expressed as an integral over all radiation modes. In this form, denoting by $A_k$ and $A(\varepsilon)$ the guided and the radiation mode amplitudes, respectively, the total power of the radiated field in the waveguide axis direction is given by \cite{Sny83}:
\be
P_{rad} = \sum_k \vert A_k \vert^2 + \int_{\varepsilon \in \mathcal{I}_{\varepsilon}} d\varepsilon \, \vert A(\varepsilon) \vert^2.
\label{expand}
\ee
On the other hand, as we have seen, resonances are scattering states of complex eigen-energy that spend definite intervals of time in the interaction zone. So resonances are not connected to radiation modes because these last do not belong to complex eigenvalues. Next we shall take full advantage of the analogy indicated above to get optical models for the resonances discussed in Section~\ref{resonances}.

\subsection{Wave propagation through a dielectric homogeneous slab}
\label{well}

As an immediate application let us consider the simplest case of electromagnetic propagation
through a finite, dielectric homogeneous material. It is convenient to rewrite the eigenvalue equation (\ref{Hel2}) as follows
\be
\left[-\left( \frac{k_0^{-2}}{2n_0} \right) \frac{d^2}{dx^2} + 1-n(x) \right] \varphi =(\varepsilon +1) \varphi.
\label{Hel3}
\ee
Considering a combination of media distributed along the $x$-axis such that there is vacuum in $(-\infty, -a) \cup (a,+\infty)$, and a medium of constant refractive index $n=U_0>1$ in $x\in [-a,a]$, the equation (\ref{Hel3}) is in correspondence with the eigenvalue equation defined by a  linear square well of depth $V_0 =\vert 1-U_0 \vert$ and width $2a$. The solutions to this equation are well known and can be found elsewhere, thought the analysis of resonances and the resolution of the transmission coefficient in terms of the FBW distribution, see Eq.~(\ref{FBW1}), can be consulted in Ref.~\cite{Fer08a}. Hereafter we shall label the regions of the media (vacuum, medium, vacuum)  from left to right as $I$, $II$ and $III$.

\section{Leaky modes and resonances}
\label{leaky}

Besides the guided and radiation modes discussed in the previous section, it is well known the existence of solutions to the Helmholtz equation with complex propagation constants that have an ``unusual behavior'': they increase exponentially in the transverse direction and suffer attenuation in the longitudinal one \cite{Sny74}. These solutions are called leaky modes and have attracted much attention over the time. The leaky modes do not belong to the set of guided and radiation modes of infinite transverse waveguides, though they constitute a discrete set of modes that can be used to approximate the continuous mode contribution to the radiation field, in regions close to the core and far from the source, in order to avoid integration over all radiation modes \cite{Hu09,Sny83,Tre14,Sam76,Zhu15b}. Yet, leaky modes have properties that are very similar to the ones of guided modes, but the former modes loss energy as they propagate along the optical axis \cite{Hu09}. In slab waveguides this leakage is due to refraction at the waveguide core-clad boundaries (refracting leaky modes \cite{Sny83}). In waveguides with curved cross-section the leakage can be also produced by the partial reflection at the boundaries due only to the curvature of the cross-section (tunneling leaky modes) \cite{Sny83,Sny74}. It can be shown that the leakage is slower in the tunneling modes compared to the refractive modes \cite{Sny74}. In any case, these modes behave much like guided modes since they are able to propagate long distances inside the guide before their power loss started to be significant. Different methods have been applied to determine the leaky modes of planar \cite{Hu09,Pet02} as well as cylindric \cite{Sny74,Sam76} waveguides involving isotropic and anisotropic dielectrics \cite{Liu14}. These techniques include numerical methods as mode-matching \cite{Ogu83}, finite differences and finite element \cite{Tre14,Ura05,Liu14}, approximate methods as WKB \cite{Gha85}, transfer matrix \cite{Pet02} and differential transfer matrix methods \cite{Zhu11,Zhu15b}. There is a conceptual discussion concerning  leaky modes, the exponential growing in the transversal direction of these nearly guided modes make them appear as nonphysical solutions to the Helmholtz equation. Yet, this behavior is required if the energy flux must be preserved: As the flux is damped in the longitudinal direction, there must be an increasing flux in the transverse direction \cite{Hu09}.

The problem of finding leaky modes is mathematically equivalent to determine the resonant states in quantum mechanics for a particle in a short range interaction potential. This is convenient because we may apply different methods that have been addressed to obtain the resonance energies and the corresponding wave functions in order to determine the complex propagation constants and the modal fields. To this end, let us regard the square well potential analogy indicated in Section~\ref{well} with a complex propagation constant, $\varepsilon = \varepsilon_R  -i\frac{\Gamma}{2}$. The corresponding complex wave number $k=k_R+ik_I$ is related to $\varepsilon$ through
\begin{equation}\label{complexk}
    \varepsilon_R = \frac{k_R^2-k_I^2}{2k_0^2}-1,\quad  -\frac{\Gamma}{2} = \frac{k_Rk_I}{k_0^2}.
\end{equation}
The energy density flux in the transverse direction is not null because $k_R\neq 0$ (see \cite{Fer08a}). A wave coming from region $III$ enters into resonance with the optical medium and is thus trapped inside the core (region $II$), then it propagates and travels a finite distance along the optical axis. After a while, the wave is finally emitted into the clad. The result is a wave propagating to the left in region $I$, and another one propagating to the right in region $III$. The correct sign for $k_R$ is thus positive and the whole process obey the pure outgoing condition. The expression for the electric energy density fulfills
\begin{equation}\label{densitycomplex}
    \lim_{x\rightarrow\pm\infty}\rho_e \propto
    e^{-k_0\Gamma z}\left| \varphi (x)\right|^2,
\end{equation}
meaning that the power decreases exponentially as $z$ grows whenever $\Gamma>0$ and $k_I<0$. The optical-quantum analogy requires also a purely outgoing wave $\varphi(x)$, see Section~\ref{resonances}, so the amplitudes of the external electric fields increase exponentially as $x\rightarrow \pm \infty$. Leaky modes then correspond to wave numbers $k=k_R +ik_I$ in the fourth quadrant of the complex plane. The factor $k_0^{-1}/\Gamma$ can be interpreted as the mean distance traveled by the beam, along the optical axis, before its intensity decays by $1/e$ of its intensity at a focus.  In this context $\Gamma$ represents a measure of the transient confinement of the field into the core. In the limit $\Gamma\rightarrow 0$ we will have a quasi-stationary mode propagating inside the slab in a very similar way as a guided mode. In this limit, it is possible to establish an analytic approach to calculate the propagation constant. Bearing this in mind, following (\ref{cond1}), we may write
\begin{equation}\label{longlife}
  \frac{\Gamma/2}{\Delta \varepsilon_R} \ll 1,
\end{equation}
where $\Delta \varepsilon_R$ is the separation between two consecutive resonances. To get the position $\varepsilon_R$ and the width $\Gamma$ of the resonances we follow the approximations reported in \cite{Fer08a}. After some calculations one arrives at the expressions
\begin{equation}\label{varep-quant}
  \varepsilon_R \approx \frac{1}{2k_0^2 U_0}\left(\frac{m\pi}{2a}\right)^2 - U_0
\end{equation}
and
\begin{equation}\label{Gamma}
  \frac{\Gamma}{2} \approx\frac{1}{k_0 a U_0}\sqrt{2\left(\varepsilon_R+1\right)},
\end{equation}
with $m$ an integer number fulfilling
\begin{equation}\label{m-values}
  \sqrt{2U_0\left(U_0-1\right)}< \frac{m\pi}{2k_0a}<\sqrt{2U_0^2}.
\end{equation}

\begin{figure}
\centering\includegraphics[width=12cm]{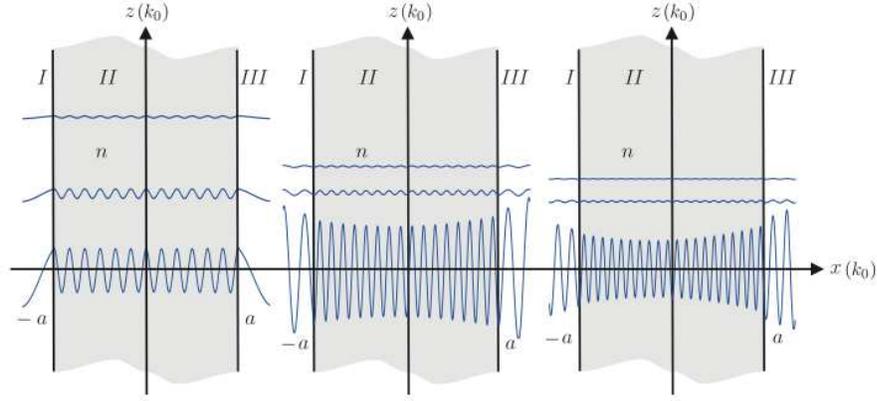}
\caption{\footnotesize Propagation of resonances inside a slab waveguide of width $k_0a=30$ and constant refractive index $n = U_0 =1.5$. The slab supports $17$ leaky modes ($24\leq m \leq 40$) defined by Equations~(\ref{varep-quant}) and (\ref{Gamma}). The figure shows the propagation in the axial direction (from the bottom to the top in each slab) of the real part of the leaky modes field amplitudes for $m=24$ (left), $m=32$ (center) and $m=40$ (right). Note that these are attenuated as the mode propagates. The leakage is faster for larger values of $m$.}
\label{res}
\end{figure}

\noindent
Thus, for each combination of the parameters $\left(a,U_0,k_0\right)$, there is a finite number of resonances given by (\ref{varep-quant}) and (\ref{Gamma}). Table \ref{tabla1} shows the eigenvalues $\varepsilon$ corresponding to the $17$ leaky modes supported by a slab waveguide of refractive index $U_0=1.5$ and width $k_0a=30$. The behavior of the real part of the field amplitude of these modes, as they propagate in the longitudinal direction through the waveguide, is depicted in Figure \ref{res} for $m=24$ (left), $m=32$ (center) and $m=40$ (right). Observe that the modes with smaller values of $m$ attenuate slower compared to those with larger values of $m$.

\begin{table}[h]
\centering
\begin{tabular}{|c c|c c|}
\hline\hline
\multicolumn{4}{|c|}{$k_0a = 30, \quad U_0=1.5$}\\[0.5ex]
\hline\hline
$m$ & $\varepsilon = \varepsilon_R - i\frac{\Gamma}{2}$ & $m$ & $\varepsilon = \varepsilon_R - i\frac{\Gamma}{2}$\\[0.5ex]
\hline
24 & $-0.973621 - i0.0051042 $ & 33 & $-0.504815 - i0.0221150 $\\
25 & $-0.928842 - i0.0083833 $ & 34 & $-0.443587 - i0.0234424 $\\
26 & $-0.882236 - i0.0107847 $ & 35 & $-0.380531 - i0.0247350 $\\
27 & $-0.833802 - i0.0128120 $ & 36 & $-0.315647 - i0.0259981 $\\
28 & $-0.783540 - i0.0146215 $ & 37 & $-0.248936 - i0.0272358 $\\
29 & $-0.731450 - i0.0162860 $ & 38 & $-0.180397 - i0.0284514 $\\
30 & $-0.677533 - i0.0178462 $ & 39 & $-0.110031 - i0.0296476 $\\
31 & $-0.621788 - i0.0193273 $ & 40 & $-0.037836 - i0.0308267 $\\
32 & $-0.564215 - i0.0207462 $ & & \\[0.5ex]
\hline\hline
\end{tabular}
\caption{\footnotesize Eigenvalues $\varepsilon$ corresponding to the $17$ leaky modes supported by a slab
of width $k_0a=30$ and constant refractive index $U_0=1.5$.}
\label{tabla1}
\end{table}

\section{Longitudinal shifts}
\label{shift}

Let us determine the longitudinal shift of a wave packet under scattering by an optical medium. This will provide us some insight of the distance traveled by the leaky mode in the waveguide before the longitudinal power is completely lost. This problem is equivalent to evaluate the reflection and transmission phase times for a particle in a rectangular well potential  \cite{Har62,Li00,Ste94}. Consider a wave-packet scattering solution for the external electric field
\begin{equation}\label{externalscat-}
    E_I(x,z) = \int_0^\kappa dk f(k) e^{i(k x-k_0\varepsilon(k) z)} + \int_0^{\kappa} dk f(k)r(k) e^{i(-k x-\varepsilon(k) k_0 z-2ka+\phi(k))},
\end{equation}
\begin{equation}\label{externalscat+}
    E_{III}(x,z) = \int_0^{\kappa}dk f(k)t(k) e^{i(k x-\varepsilon(k)k_0 z -2ka +\phi(k) +\pi/2)},
\end{equation}
where $f(k)$ is the Fourier coefficient, $\kappa$ is defined by $\varepsilon(\kappa)=0$, $r(k)$ and $t(k)$ are the reflexion and transmission amplitudes respectively and
\begin{equation}\label{phase}
  \phi(k) = -\arctan \frac{2kq\cos 2qa}{\left(k^2+q^2\right)\sin 2qa}, \quad q^2 = U_0\left(k^2 + 2k_0^2 \left(U_0-1\right)\right).
\end{equation}

\begin{figure}[h]
\centering\includegraphics[width=14cm]{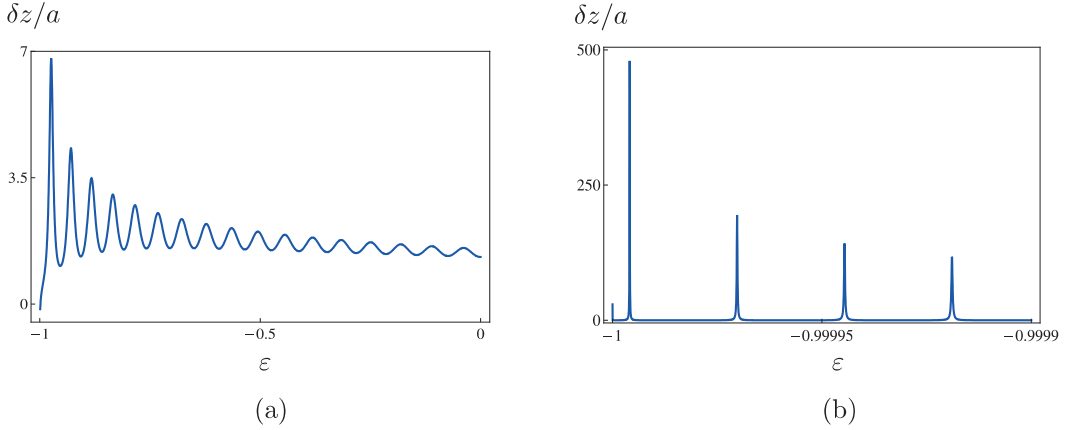}
\caption{\footnotesize Longitudinal shift as a function of the eigenvalue $\varepsilon = -p_z$ for a dielectric slab of width $k_0a$ equal to (a) $30$ and (b) $50,000$. In both cases $U_0=1.5$.}
\label{delay-en}
\end{figure}

\noindent
Applying the stationary phase condition \cite{Har62,Ros12,Ros13} to the incident and transmitted wave packets one has
\begin{equation}\label{phase-inc}
  \frac{\partial}{\partial k} \left(kx -\frac{k^2}{2k_0}z + k_0z\right)=0, \quad \frac{\partial}{\partial k} \left(kx-2ka-\frac{k^2}{2k_0}z + k_0z + \phi(k)+ \frac{\pi}{2}\right)=0.
\end{equation}
So that the equations of motion for the centers of the corresponding Fourier components are given by
\begin{equation}\label{trans-eq}
  x = \frac{k}{k_0}z, \quad x = 2a +\frac{k}{k_0}z + \frac{\partial \phi}{\partial k}.
\end{equation}
The point $z_{in}$ at which an incident plane wave of wave number $k$ would reach the left wall of the optical medium is then $z_{in}=-\frac{k_0}{k}a$. In the same way the point $z_t$ at which the reflected or transmitted wave leaves the optical medium is
\begin{equation}\label{zref-tr}
  z_t=\frac{k_0}{k} \left(-a+ \frac{\partial \phi}{\partial k}\right).
\end{equation}
Then the longitudinal shift of the reflected or transmitted wave is readily written
\begin{equation}\label{deltaz}
  \delta z =z_t - z_{in}= \frac{k_0}{k} \frac{\partial \phi}{\partial k}.
\end{equation}

\begin{figure}[h]
\centering\includegraphics[width=8cm]{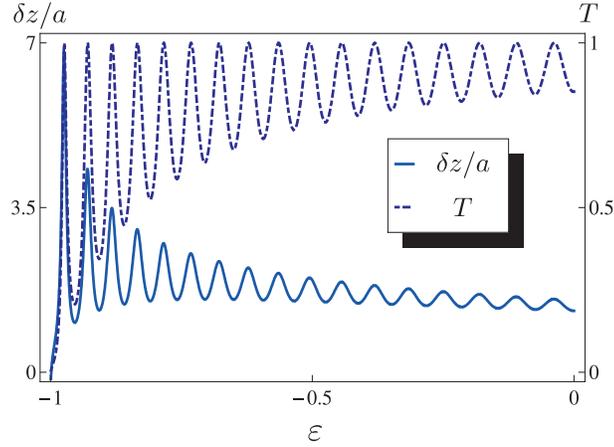}
\caption{\footnotesize Longitudinal shift (continuous line) and transmission coefficient (dotted line) of an electric wave propagating in a dielectric slab of width $k_0 a= 30$, and constant refractive index $U_0=1.5$, both as a function of the eigenvalue $\varepsilon$. Observe that the maxima of both curves coincide.}
\label{delay-trans}
\end{figure}

\begin{figure}[h]
\centering\includegraphics[width=7cm]{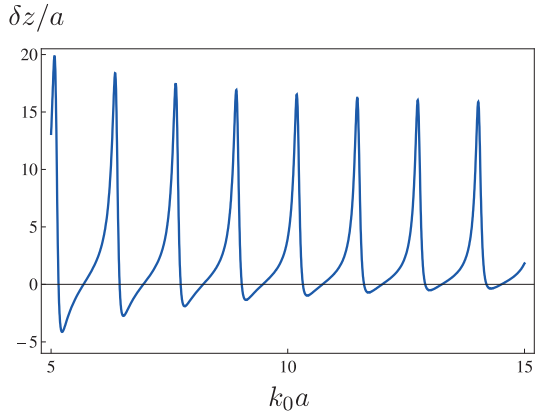}
\caption{\footnotesize (a) Longitudinal shift as a function of the width $k_0a$ of a dielectric slab with constant refractive index $U_0 =1.5$. Here $\varepsilon = -0.995$.}
\label{delay-wi}
\end{figure}

\noindent
Figure \ref{delay-en} shows the longitudinal shift of a Fourier component under reflection or transmission as a function of $\varepsilon$ for (a) $k_0a=30$ and (b) $k_0a=50,000$. It is worthwhile to point out that the peaks of this shift coincide with the peaks of the corresponding transmission coefficient as shown in Figure \ref{delay-trans}. This implies that the electromagnetic modes having maximum transmission probabilities, and thus corresponding to the resonant (leaky) modes, travel a maximum longitudinal distance in the core before they are finally transmitted to the infinite clad. Figure \ref{delay-wi} shows the longitudinal shift as a function of the width $k_0a$ for the eigenvalue $\varepsilon = -0.995$ and $U_0=1.5$. We can appreciate  that the transmission shift $\delta z$ takes negative values, these have been already predicted for both rectangular barriers and wells in the quantum approach \cite{Li00,Ste94}. The analogy between the Schr\"odinger and the paraxial Helmholtz equations suggest that this phenomenon may be used to measure the negative phase time in the scattering of particles by a potential well (compare with \cite{Vet01}).

\section{Conclusions}
\label{concluye}

The classical optics analogy of quantum systems is based on the relationship between the Helmholtz equation and the two degrees of freedom Schr\"odinger equation. Although such analogy is mathematical, the optical solutions are useful in simulating the quantum ones for diverse cases. For instance, it is well known that the guided and radiation modes of a waveguide structure can be put in correspondence with the bound and scattering states of a one-dimensional short range potential in quantum mechanics. However, the association of leaky modes with their quantum counterpart, as well as the identification of the optical analogue of a quantum resonance, is rarely discussed in the literature on the matter. One of the reasons would be that in both cases the calculation of the eigenvalues and wave-solutions is not as immediate as in the scattering (radiation) or bound (guided) states (modes). In this work we have presented a form to connect these two kinds of wave solutions: a leaky mode is in correspondence with a resonance state. The former is as close to a guided mode as a resonance is close to a bound state. That is, a leaky mode (resonance state) is a guided mode (bound state) for which the longitudinal shift (lifetime) is definite.

\section*{Acknowledgments}
The financial support of CONACyT and Project SIP20150200 (Instituto Polit\'ecnico Nacional) is acknowledged.


\end{document}